\documentclass[journal]{IEEEtran}
\IEEEoverridecommandlockouts

\usepackage{cite}
\usepackage{amsmath,amssymb,amsfonts}
\usepackage{algorithmic}
\usepackage{graphicx}
\usepackage{textcomp}
\usepackage{xcolor}
\usepackage{array}
\usepackage{caption}
\usepackage{booktabs}
\usepackage{multirow}
\usepackage{soul}
\usepackage{subcaption}
\usepackage{balance}
\usepackage[capitalise]{cleveref}
\crefname{section}{Sec.}{Secs.}
\crefname{figure}{Fig.}{Figs.}
\usepackage{tikz}
\usetikzlibrary{fit}
\makeatletter
\tikzset{
  fitting node/.style={
    inner sep=0pt,
    fill=none,
    draw=none,
    reset transform,
    fit={(\pgf@pathminx,\pgf@pathminy) (\pgf@pathmaxx,\pgf@pathmaxy)}
  },
  reset transform/.code={\pgftransformreset}
}
\makeatother

\usepackage{pgfplots}
\pgfplotsset{compat=newest} 
\pgfplotsset{plot coordinates/math parser=false} 

\usepackage[acronyms,nonumberlist,nopostdot,nomain,nogroupskip,acronymlists={hidden}]{glossaries}
\newlength\fheight
\newlength\fwidth
\def\BibTeX{{\rm B\kern-.05em{\sc i\kern-.025em b}\kern-.08em
    T\kern-.1667em\lower.7ex\hbox{E}\kern-.125emX}}

\usepackage{tikzpagenodes,etoolbox}
\usetikzlibrary{calc}
\usepackage[contents={}]{background}

\AddEverypageHook{%
  \ifnumequal{\value{page}}{1}{%
    \tikz[remember picture,overlay]{%
      \node[
        draw,
        minimum width=1.03\textwidth,
        text width=1.02\textwidth,
        font=\scriptsize,
        align=center,
        inner sep=4pt
      ]
      at ($(current page header area) - (0,5pt)$)
      {%
        This work has been submitted to the IEEE for possible publication.\\
        Copyright may be transferred without notice, after which this version may no longer be accessible.
      };
    }%
  }{}%
}

\title{Toward Native ISAC Support in O-RAN Architectures for 6G}

\author{
    \IEEEauthorblockN{
        Eduardo Baena\IEEEauthorrefmark{1},
        Rajesh Krishnan\IEEEauthorrefmark{1}, 
        Mai Vu\IEEEauthorrefmark{2}, 
        Gil Zussman\IEEEauthorrefmark{3},
        Dimitrios Koutsonikolas\IEEEauthorrefmark{1}, 
    }

    \IEEEauthorblockA{
        \IEEEauthorrefmark{1}Institute for the Wireless Internet of Things, Northeastern University, Boston, MA, USA\\
        \IEEEauthorrefmark{2}Tufts University, Medford, MA, USA,\\
        \IEEEauthorrefmark{3}Columbia University, New York, NY, USA
    }

}

\IEEEoverridecommandlockouts

\begin{document}
\newacronym{3gpp}{3GPP}{Third Generation Partnership Project}
\newacronym{5g}{5G}{Fifth-Generation}
\newacronym{6g}{6G}{Sixth-Generation}
\newacronym{a1}{A1}{O-RAN Policy Management Interface}
\newacronym{ai}{AI}{Artificial Intelligence}
\newacronym{awgn}{AWGN}{Average White Gaussian Noise}
\newacronym{cpu}{CPU}{Central Processing Unit}
\newacronym{cu}{CU}{Centralized Unit}
\newacronym{daa}{DAA}{Decentralized Authentication Authority}
\newacronym{dapp}{dApp}{Distributed RAN Application}
\newacronym{dt}{DT}{Digital Twin}
\newacronym{du}{DU}{Distributed Unit}
\newacronym{e2}{E2}{O-RAN Near-Real-Time Interface}
\newacronym{ecdf}{ECDF}{Empirical Cumulative Distribution Function}
\newacronym{fede2}{FED-E2}{Federated O-RAN Near-Real-Time Interface}
\newacronym{fl}{FL}{Feeder Link}
\newacronym{geo}{GEO}{Geostationary Orbit}
\newacronym{gpu}{GPU}{Graphics Processing Unit}
\newacronym{gs}{GS}{Geometric Spread}
\newacronym{hap}{HAP}{High Altitude Platform}
\newacronym{hdtn}{HDTN}{High-Delay Tolerant Networking}
\newacronym{isac}{ISAC}{Integrated Sensing and Communication}
\newacronym{iq}{IQ}{In-phase and Quadrature}
\newacronym{iab}{IAB}{Integrated Access Backhaul}
\newacronym{isl}{ISL}{Inter-Satellite Link}
\newacronym{leo}{LEO}{Low-Earth Orbit}
\newacronym{los}{LOS}{Line-Of-Sight}
\newacronym{meo}{MEO}{Medium Earth Orbit}
\newacronym{mec}{MEC}{Mobile Edge Computing}
\newacronym{ml}{ML}{Machine Learning}
\newacronym{mmwave}{mmWave}{millimeter wave}
\newacronym{mppt}{MPPT}{ Maximum Power Point Tracking }
\newacronym{near-rt-ric}{near-RT-RIC}{near-RT-\gls{ric}}
\newacronym{nr}{NR}{New Radio}
\newacronym{ntn}{NTN}{Non-Terrestrial Network}
\newacronym{o1}{O1}{O-RAN Operations and Management Interface}

\newacronym{aoa}{AoA}{Angle of Arrival}
\newacronym{fd}{FD}{Full Duplex}
\newacronym{fft}{FFT}{Fast Fourier Transform}
\newacronym{prs}{PRS}{Positioning Reference Signal}
\newacronym{ofdm}{OFDM}{Orthogonal Frequency Division Multiplexing}
\newacronym{os}{OS}{Operating System}
\newacronym{osiran}{OSIRAN}{Open-Space-Integrated-RAN}
\newacronym{pe}{PE}{Pointing Error}
\newacronym{qos}{QoS}{Quality of Service}
\newacronym{ran}{RAN}{Radio Access Network}
\newacronym{ric}{RIC}{RAN Intelligent Controller}
\newacronym{rapp}{rApp}{RAN Application}
\newacronym{sapp}{sApp}{Space Application}
\newacronym{sagin}{SAGIN}{Space-Air-Ground Integrated Network}
\newacronym{sl}{SL}{Service Link}
\newacronym{smo}{SMO}{Service Management and Orchestration}
\newacronym{spaceran}{Space-RAN}{Space integrated Open Radio Access network}
\newacronym{spaceric}{Space-RIC}{Space RAN Intelligent Controller}
\newacronym{sru}{s-RU}{Satellite Radio Unit}
\newacronym{sdu}{s-DU}{Satellite Distributed Unit}
\newacronym{scu}{s-CU}{Satellite Central Unit}
\newacronym{snf}{s-NF}{Satellite Network Function}
\newacronym{tn}{TN}{Terrestrial Network}
\newacronym{ue}{UE}{User Equipment}
\newacronym{ul}{UL}{User Link}
\newacronym{xapp}{xApp}{eXtended RAN Application}
\newacronym{gsl}{GSL}{ground-to-satellite links}
\newacronym{kpi}{KPI}{Key Performance Indicator}
\newacronym{sic}{SIC}{Self-Interference Cancellation}
\newacronym{srs}{SRS}{Sounding Reference Signal}
\newacronym{mimo}{MIMO}{Multiple-Input Multiple-Output}
\newacronym{mumimo}{MU-MIMO}{Multi-User MIMO}
\newacronym{paam}{PAAM}{Phased Array Antenna Module}
\newacronym{ofh}{O-FH}{Open Fronthaul}
\maketitle

\glsresetall

\begin{abstract}

\gls{isac} is an emerging paradigm in 6G networks that enables environmental sensing using wireless communication infrastructure. Current O-RAN specifications lack the architectural primitives for sensing integration: no service models expose physical-layer observables, no execution frameworks support sub-millisecond sensing tasks, and fronthaul interfaces cannot correlate transmitted waveforms with their reflections.

This article proposes three extensions to O-RAN for monostatic sensing, where transmission and reception are co-located at the base station. First, we specify sensing dApps at the O-DU that process IQ samples to extract delay, Doppler, and angular features. Second, we define E2SM-SENS, a service model enabling xApps to subscribe to sensing telemetry with configurable periodicity. Third, we identify required Open Fronthaul metadata for waveform-echo association. We validate the architecture through a prototype implementation using beamforming and Full-Duplex operation, demonstrating closed-loop control with median end-to-end latency suitable for near-real-time sensing applications. While focused on monostatic configurations, the proposed interfaces extend to bistatic and cooperative sensing scenarios.

\end{abstract}



\glsresetall
\glsunset{rapp}
\glsunset{xapp}
\glsunset{dapp}

\section{Introduction}

\gls{isac} combines wireless communication and environmental sensing within shared radio infrastructure, enabling 6G networks to detect objects, estimate motion, and provide spatial awareness without dedicated sensing hardware~\cite{3gpp2024, orannGRG2023}. Standardization efforts including 3GPP TR~22.837 (Release~19) and O-RAN Alliance reports have identified ISAC as a key 6G capability~\cite{etsi2024}, but current architectures lack the primitives for practical integration.

The O-RAN architecture offers a promising foundation through its disaggregated design and programmable control hierarchy. The \gls{near-rt-ric} enables policy-based orchestration at 10--100~ms timescales, while distributed RAN applications (dApps) at the \gls{du} support sub-millisecond execution~\cite{doro2025dapps}. However, three gaps prevent ISAC realization: (1) no execution frameworks at the O-DU support real-time sensing signal processing; (2) no service models expose sensing-derived metrics to the control plane; and (3) the \gls{ofh} interface lacks mechanisms to associate received IQ samples with their corresponding transmitted waveforms.

This paper addresses these gaps with three contributions:
\begin{itemize}
    \item \textbf{Sensing dApps:} We specify DU-level applications that process IQ samples from the O-FH to extract sensing features including delay, Doppler shift, and angle-of-arrival (AoA).
    \item \textbf{E2SM-SENS:} We define a new E2 service model enabling xApps to subscribe to sensing telemetry with configurable reporting periodicity, following a publisher-subscriber pattern.
    \item \textbf{O-FH metadata requirements:} We identify the timing markers and waveform descriptors needed for coherent sensing processing across the RU-DU interface.
\end{itemize}

We focus on \textit{monostatic} sensing configurations, where transmission and echo reception occur at the same base station. This approach leverages beamforming for spatial selectivity and Full-Duplex (FD) operation for simultaneous transmit-receive capability. While bistatic and cooperative sensing introduce additional coordination requirements, the proposed interfaces provide a foundation for these extensions.

We validate the architecture through a prototype implementation of E2SM-SENS with a sensing dApp that processes IQ samples using spectral analysis. Experimental results demonstrate that the system achieves configurable telemetry rates and closed-loop latencies compatible with near-real-time sensing applications.

The remainder of this paper is organized as follows. \cref{sec:background} reviews ISAC fundamentals and the O-RAN architecture. \cref{sec:challenges} identifies architectural challenges and use cases. \cref{sec:architecture} presents the proposed extensions. \cref{sec:evaluation} describes the prototype implementation and experimental results. \cref{sec:future} discusses open research directions, and \cref{sec:conclusions} concludes the paper.


\section{Background and Related Work}
\label{sec:background}

\subsection{ISAC Sensing Configurations}

ISAC systems embed radar-like functionality into communication waveforms, particularly \gls{ofdm}, enabling networks to estimate target position, velocity, and trajectory from signal reflections~\cite{liu2022, wild2021}. The sensing configuration depends on the spatial relationship between transmitter and receiver:

\begin{itemize}
    \item \textbf{Monostatic:} Transmission and reception occur at the same node, analogous to traditional radar. This requires simultaneous transmit-receive capability, typically achieved through \gls{fd} operation with \gls{sic}. In beamformed systems, sensing can operate in two modes: dedicated probing beams that illuminate regions of interest independently of user traffic, or opportunistic sensing that extracts environmental information from communication beam reflections.
    \item \textbf{Bistatic:} Transmitter and receiver are spatially separated, requiring synchronization between nodes but avoiding self-interference challenges.
  
    \item \textbf{Multistatic/Cooperative:} Multiple nodes coordinate sensing operations, enabling extended coverage and improved accuracy through spatial diversity.
\end{itemize}

This work focuses on monostatic sensing, where beamforming provides spatial selectivity and FD enables echo reception during transmission. The architectural extensions we propose, particularly E2SM-SENS telemetry and O-FH metadata, provide foundations that extend to bistatic configurations through inter-node coordination via the Near-RT RIC.

\subsection{O-RAN Architecture Overview}

The O-RAN architecture disaggregates the RAN into functional components connected through open interfaces. The \gls{du} performs baseband processing including channel coding and resource mapping. The Radio Unit (O-RU) handles RF transmission and reception. The \gls{near-rt-ric} hosts xApps that implement control logic at 10--100~ms timescales, while the Non-RT RIC hosts rApps for policy-based management at longer timescales.

Recent work has introduced dApps, distributed applications that execute directly at the O-DU with sub-millisecond latency~\cite{doro2025dapps}. Unlike xApps, which receive abstracted metrics over the E2 interface, dApps can access IQ samples and PHY-layer state, enabling real-time signal processing tasks. Fig.~\ref{fig:control_hierarchy} illustrates this layered architecture and the interfaces connecting each component.

\subsection{Related Work}

ISAC standardization is progressing across multiple bodies. 3GPP Release~19 delivered TR~22.837 (32 use cases across five scenarios) and TS~22.137 defining key sensing requirements, including positioning (1--10~m) and velocity accuracy (1--10~m/s)~\cite{3gpp2024}. Release~20 initiates network-based sensing studies, while normative 6G ISAC specifications are expected in Release~21 with ASN.1/OpenAPI freeze in March~2029. In parallel, O-RAN nGRG reports identify sub-millisecond control loops as essential for radar processing~\cite{orannGRG2023}, and ETSI ISG ISAC published use-case and channel-modeling reports with formal KPIs in 2025~\cite{etsi2024}. However, standardized interfaces for sensing telemetry and fronthaul metadata remain undefined.

Hamidi-Sepehr et al.~\cite{hamidi2024} demonstrate that fronthaul capacity can support sensing data streams under Split~7.2. However, their analysis assumes ideal DU processing and omits system-level concerns including task prioritization, scheduling conflicts, and integration of sensing outputs into control logic. Recent work on dApps~\cite{doro2025dapps} demonstrates real-time positioning using channel impulse response extraction, validating the feasibility of sub-millisecond sensing at the DU. Villa et al.~\cite{villa2025gpu} extend this direction with GPU-accelerated dApps on commercial gNB hardware, achieving sub-2 ms inference latency for CSI-based localization; however, their UL-collaborative approach requires an active UE transmission, whereas our monostatic architecture enables autonomous environmental sensing through dedicated probing beams without UE involvement. Ludant et al.~\cite{ludant2025isac_oran} analyze radar processing placement tradeoffs in OpenRAN, highlighting security implications of fronthaul IQ exposure. Our work addresses these gaps through concrete architectural extensions and experimental validation of closed-loop latency.

\section{Challenges and Use Cases}
\label{sec:challenges}

\subsection{Challenges}

Current O-RAN specifications present architectural gaps that prevent practical ISAC integration. The fundamental challenge is temporal: sensing operations require sub-millisecond feedback loops to maintain spatial coherence during target tracking, while the Near-RT RIC operates at 10--100~ms cycles~\cite{polese:jsac2024}. This order-of-magnitude timing mismatch means that sensing-derived insights arrive too late to influence beam steering or interference management decisions, necessitating execution at the O-DU level through dApps.

Beyond timing constraints, three structural gaps impede integration: (1) existing E2 service models (KPM, RC, NI) address communication-centric metrics exclusively, with no abstractions for sensing observables such as range, Doppler, or \gls{aoa}; (2) the O-FH U-Plane lacks mechanisms to correlate received IQ samples with their corresponding transmitted waveforms, preventing coherent echo processing; and (3) no resource partitioning exists between sensing and communication tasks at the DU, creating potential contention for baseband processing cycles and RF chain access.

\subsection{Use Cases and Requirements}

Table~\ref{tab:use_cases} presents representative ISAC use cases with their sensing requirements. Each application domain imposes distinct constraints on the underlying architecture:

\textbf{Vehicular perception:} Intelligent transportation systems require meter-level localization and collaborative perception in rapidly changing environments. Infrastructure-based sensing can detect occluded objects through indirect RF reflections, complementing onboard sensors where GNSS signals are degraded. The proposed dApp architecture enables real-time fusion of sensing data with communication scheduling, supporting lane-level awareness with $<$10~ms feedback loops.

\textbf{UAV tracking and coordination:} Unmanned aerial operations in congested airspace demand precise Doppler and angular estimation for collision avoidance. RF-based sensing provides passive trajectory estimation without requiring cooperative transponders. The E2SM-SENS telemetry model enables xApps to subscribe to UAV-specific KPIs and trigger adaptive beam tracking when trajectory changes are detected.

\textbf{Industrial automation:} Factory environments with mobile robots impose the most stringent latency requirements ($<$1~ms) due to closed-loop control constraints. Severe multipath and interference in industrial settings require robust sensing algorithms executed locally at the DU through specialized dApps, avoiding the latency penalties of centralized processing.

\textbf{Predictive beam management:} Spatial awareness from environmental sensing enables proactive reconfiguration before channel degradation occurs. Detecting obstacles or user trajectory changes through sensing feedback allows the system to prepare handover or beam switching decisions, reducing reactive latency and improving link reliability.

\begin{table}[h]
\centering
\small
\caption{ISAC Use Cases and Sensing Requirements~\cite{3gpp2024, etsi2024}}
\label{tab:use_cases}
\begin{tabular}{@{}lll@{}}
\toprule
\textbf{Use Case} & \textbf{Key KPIs} & \textbf{Latency} \\
\midrule
Vehicular perception & Range 1--10~m, Vel.\ 1--10~m/s & $<$10~ms \\
UAV tracking         & Position 1--10~m, Vel.\ 1--10~m/s & $<$20~ms \\
Industrial control   & Position $<$1~m               & $<$1~ms \\
Beam management      & Angular update rate            & $<$5~ms \\
\bottomrule
\end{tabular}
\end{table}

These requirements motivate the architectural extensions presented in the following section.

\section{Architectural Integration of ISAC in O-RAN}
\label{sec:architecture}

This section presents architectural extensions addressing the gaps identified above: (1) dApps for real-time sensing at the O-DU; (2) E2SM-SENS for control-plane telemetry; (3) O-FH metadata for waveform correlation; and (4) multi-layer control integration.

\subsection{Real-Time Execution at the DU: dApps for ISAC}

ISAC functionalities impose latency requirements that cannot be satisfied by conventional control loops. Operations that extract physical-layer information from incoming signals must respond within sub-millisecond timescales, especially under high mobility, dense multipath, or fast-changing RF conditions. These temporal constraints exceed the control periodicity and transport latencies inherent to Near-RT RIC-based decision making~\cite{polese:jsac2024}.

To address this challenge, the proposed architecture introduces a sensing processing dApp at the O-DU. Unlike xApps, which operate asynchronously over abstracted metrics, dApps~\cite{doro2025dapps} can process IQ samples directly at the radio edge, without intermediate buffering or protocol indirection. This direct access allows time-critical signal features to be extracted in situ and supports immediate adaptation of transmission or reception parameters.

The execution pipeline is structured around lightweight routines optimized for low-latency environments. These may include temporal correlation, spectral analysis, or spatial filtering operations applied to received waveforms. When executed locally at the DU, these routines allow the system to detect relevant physical events (such as motion, reflectors, or sudden changes in propagation paths) and react within a single slot duration.

By containing sensing logic within the DU, the system preserves real-time responsiveness and avoids the need for cross-layer coordination during critical feedback cycles. This local execution also reduces congestion over control interfaces and maintains compatibility with the functional split defined in O-RAN. The output of these routines forms the basis for structured sensing metrics, which are handled by the control plane and described in the next subsection.

This encapsulation of sensing functions as dApps represents a key architectural contribution. First, it enables software-defined ISAC: sensing algorithms can be deployed, updated, or replaced without hardware modifications, accelerating innovation cycles. Second, it leverages the existing O-RAN ecosystem, where dApps, xApps, and rApps collaborate across timescales. Sensing dApps handle microsecond-level signal processing, ISAC xApps orchestrate resource allocation at millisecond granularity, and rApps define strategic sensing policies. This layered approach mirrors the proven communication control hierarchy while extending it to environmental perception. Third, the modular design supports diverse sensing applications through specialized dApps (e.g., radar processing, positioning, spectrum sensing) that share a common telemetry interface.

\begin{figure}[h]
\centering
\includegraphics[width=0.8\linewidth]{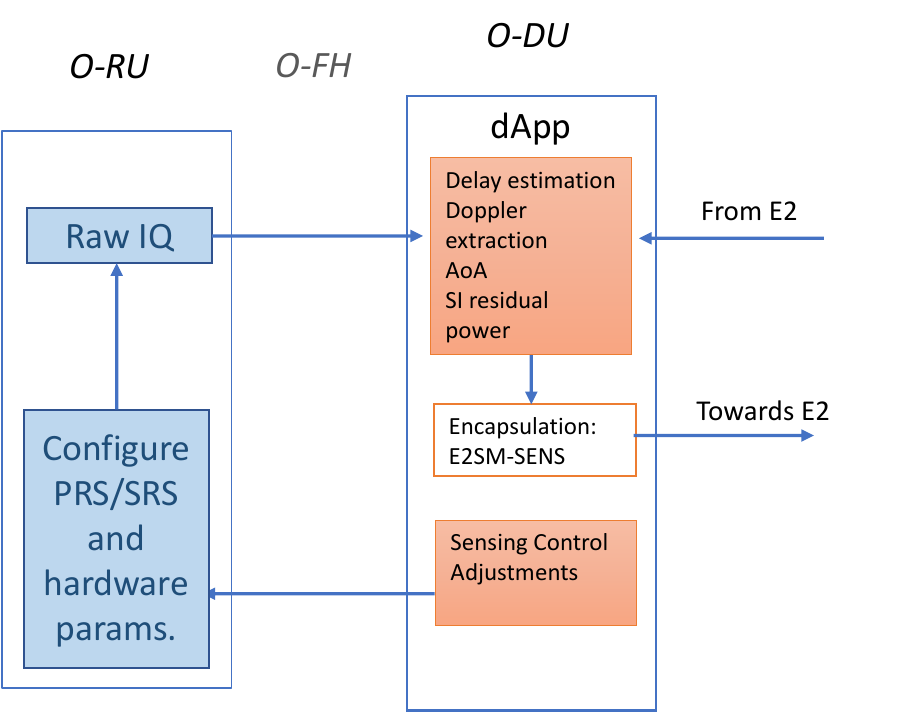}
\caption{Closed-loop sensing control via E2SM-SENS: upward KPI telemetry and downward configuration commands.}
\label{fig:e2sm_sens_flow}
\end{figure}
\begin{figure*}[h]
\centering
\includegraphics[width=0.7\linewidth]{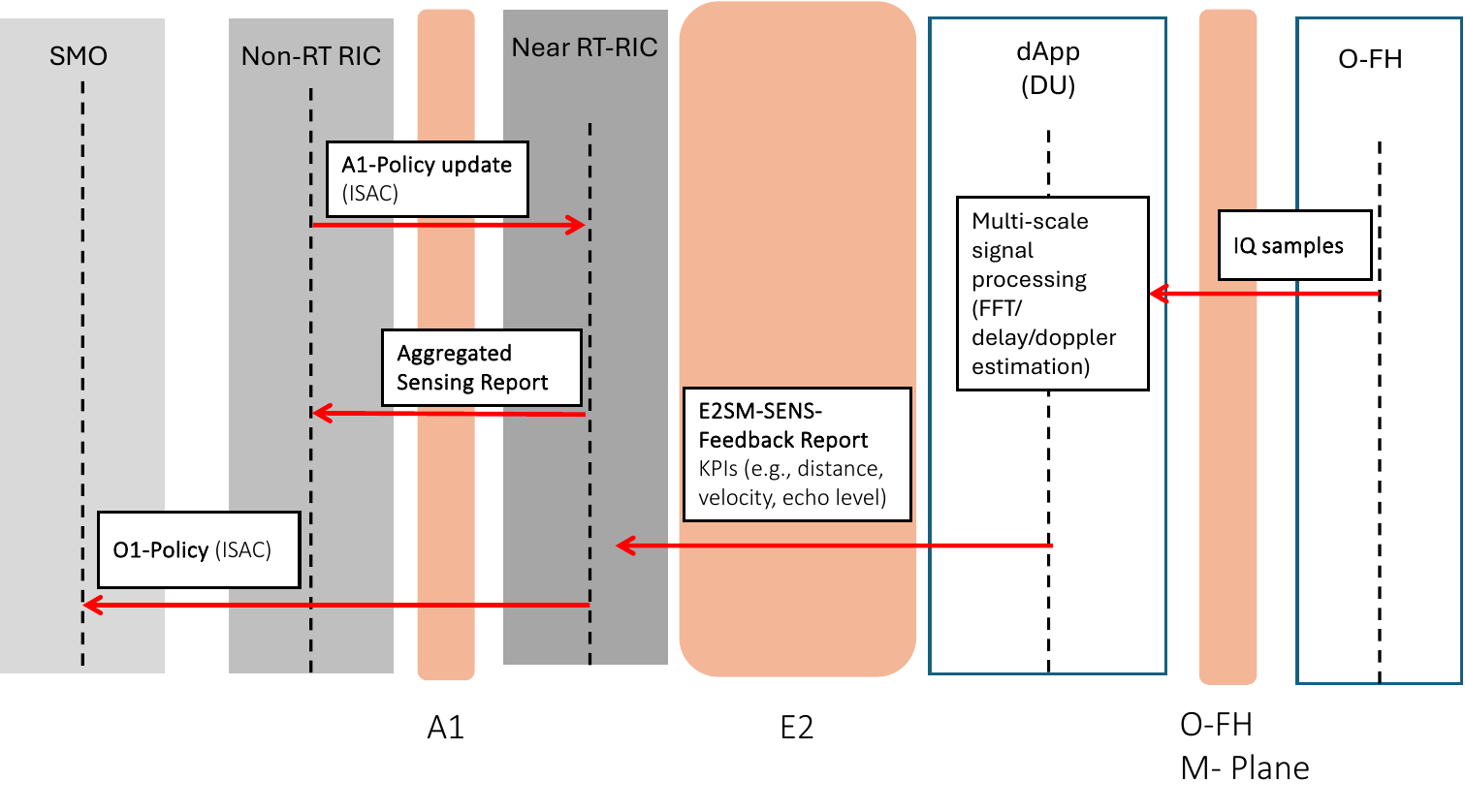}
\caption{Hierarchical ISAC control integration in O-RAN.}
\label{fig:control_hierarchy}
\end{figure*}

\subsection{Control Plane Models: E2SM-SENS and A1 ISAC Policies}

The O-RAN control plane requires the extension of existing service models to capture sensing-specific observables and manage their orchestration across time scales. Current models such as E2SM-KPM and E2SM-RC are limited to communication-centric metrics and do not expose the physical-layer measurements relevant to sensing.

To address this gap, we propose the E2SM-SENS service model, which enables structured reporting and configuration of sensing KPIs over the E2 interface. These KPIs include metrics such as delay, Doppler shift, AoA, self-interference power, and environmental indicators (e.g., multipath spread or angular entropy). These metrics are processed locally by a sensing dApp and transmitted to the control plane, where they are tagged with timestamps and confidence levels for more precise and reliable reporting.

Telemetry subscriptions follow standard E2 procedures: xApps can request periodic updates (e.g., every 5 ms) or event-driven triggers based on conditions such as threshold crossings in echo energy or sudden shifts in AoA. This allows xApps to maintain a consistent representation of the sensing context and issue E2 control messages to adapt beam configurations, sensing intervals, or self-interference canceler states as needed~\cite{polese:jsac2024}.

Longer-term objectives, like optimizing coverage in cluttered environments or allocating sensing power under spectral constraints, are encoded via A1 policies. We define a new class, \texttt{A1\_POLICY\_TYPE\_ISAC}, through which rApps specify high-level sensing directives using attributes such as geographic scope, temporal budgets, sensing priority, or energy limits. These are transmitted to xApps for enforcement, enabling hierarchical coordination between strategic planning and real-time adaptation~\cite{3gpp2024, ORANreport:dApp}.

This division of roles, strategic policy via A1, real-time telemetry, and control via E2, preserves the modular separation of the O-RAN architecture while enabling responsive ISAC behavior. The proposed extensions align with existing interface abstractions and can be incrementally integrated without modifying the underlying communication stack.


\subsection{Data Plane and O-FH Requirements for Sensing}

Monostatic sensing requires correlating received IQ samples with transmitted waveforms to extract propagation delay and Doppler shift. Current O-FH specifications (Split~7.2x) do not provide the metadata necessary for this correlation. We identify three required extensions:

\textbf{Timing markers:} Each uplink IQ block must include a high-resolution timestamp (sub-microsecond) synchronized with the transmission time of the corresponding downlink waveform. This enables round-trip delay estimation with the precision required for sub-meter range resolution.

\textbf{Waveform identifiers:} A field in the O-FH U-Plane header must associate received samples with the specific \gls{prs} or \gls{srs} waveform that generated the echo. The waveform ID indexes a configuration table containing parameters essential for matched filtering: FFT size, cyclic prefix length, subcarrier spacing, and pilot pattern. This allows the DU to select the appropriate processing pipeline without explicit parameter transmission on each frame.

\textbf{Beam association:} For beamformed transmissions, metadata must indicate the beam index and steering direction, enabling the DU to compute \gls{aoa} from the received signal.

Table~\ref{tab:ofh_extensions} summarizes these requirements with indicative field sizes.

\begin{table}[h]
\centering
\small
\caption{Required O-FH Metadata Extensions for Sensing}
\label{tab:ofh_extensions}
\begin{tabular}{|p{2.5cm}|p{2.2cm}|p{2.5cm}|}
\hline
\textbf{Field} & \textbf{Size} & \textbf{Purpose} \\
\hline
TX timestamp & 64 bits & Round-trip delay \\
\hline
Waveform ID & 16 bits & Config table index (FFT, CP, SCS) \\
\hline
Beam index & 8 bits & AoA computation \\
\hline
Sensing flag & 1 bit & Mark sensing-enabled slots \\
\hline
\end{tabular}
\end{table}

\textbf{Functional split considerations:} Under Split~7.2, baseband processing resides at the DU, making IQ-level sensing feasible. For massive-MIMO deployments where beamforming occurs at the RU (Split~7.2x Category~B), raw IQ transport becomes impractical due to fronthaul capacity constraints: a 64-antenna system at 100~MHz bandwidth with 16-bit I/Q generates approximately 200~Gbps of raw data, exceeding typical 25~Gbps eCPRI links by an order of magnitude~\cite{hamidi2024}. In such configurations, the RU must perform local preprocessing (beam-domain projection or energy detection) and report compact sensing features rather than raw samples, reducing the sensing data to the order of the number of beams rather than antennas. This requires additional signaling to convey the structure and semantics of preprocessed data, an extension not currently defined in O-RAN specifications.

\subsection{Multi-Layer Control Integration}

Figure~\ref{fig:control_hierarchy} illustrates the hierarchical control integration. At the O-DU, dApps perform sub-millisecond sensing tasks (delay/Doppler estimation, beamforming, SIC tuning). At the Near-RT RIC, an ISAC xApp aggregates E2SM-SENS telemetry and orchestrates multi-DU coordination, managing trade-offs between sensing fidelity and communication throughput. At the Non-RT RIC, rApps define long-term policies via A1 (spatial priorities, update periodicities, cross-site objectives). This separation ensures each function executes within appropriate latency constraints.


\subsection{Conceptual Example: ISAC with MU-MIMO Beamforming and Full-Duplex}

To illustrate how the proposed architecture enables ISAC, we present a conceptual deployment using \glspl{paam} with Full-Duplex capability. In \gls{mumimo} systems, spatially multiplexed beams can serve as opportunistic sensing probes: their reflections encode object position, motion, and geometry. To capture echoes in real time without interrupting transmissions, \gls{fd} capability is required at the RU, with \gls{sic} operating on microsecond timescales. Recent advances in wideband adaptive FD radios demonstrate the feasibility of real-time self-interference cancellation across wide bandwidths~\cite{gen3_implementation}.

In this conceptual architecture, a PAAM dApp at the DU would interface with phased array hardware such as that available in testbeds like COSMOS~\cite{raychaudhuri:mobicom2020}. This dApp would manage beamforming, extract angular features, and control the FD capability embedded in the radio. When FD is enabled, an additional RF chain carries the receive path, whose IQ samples are transported via O-FH with the metadata extensions described in Table~\ref{tab:ofh_extensions}.

\begin{figure}[h]
\centering
\includegraphics[width=0.9\linewidth]{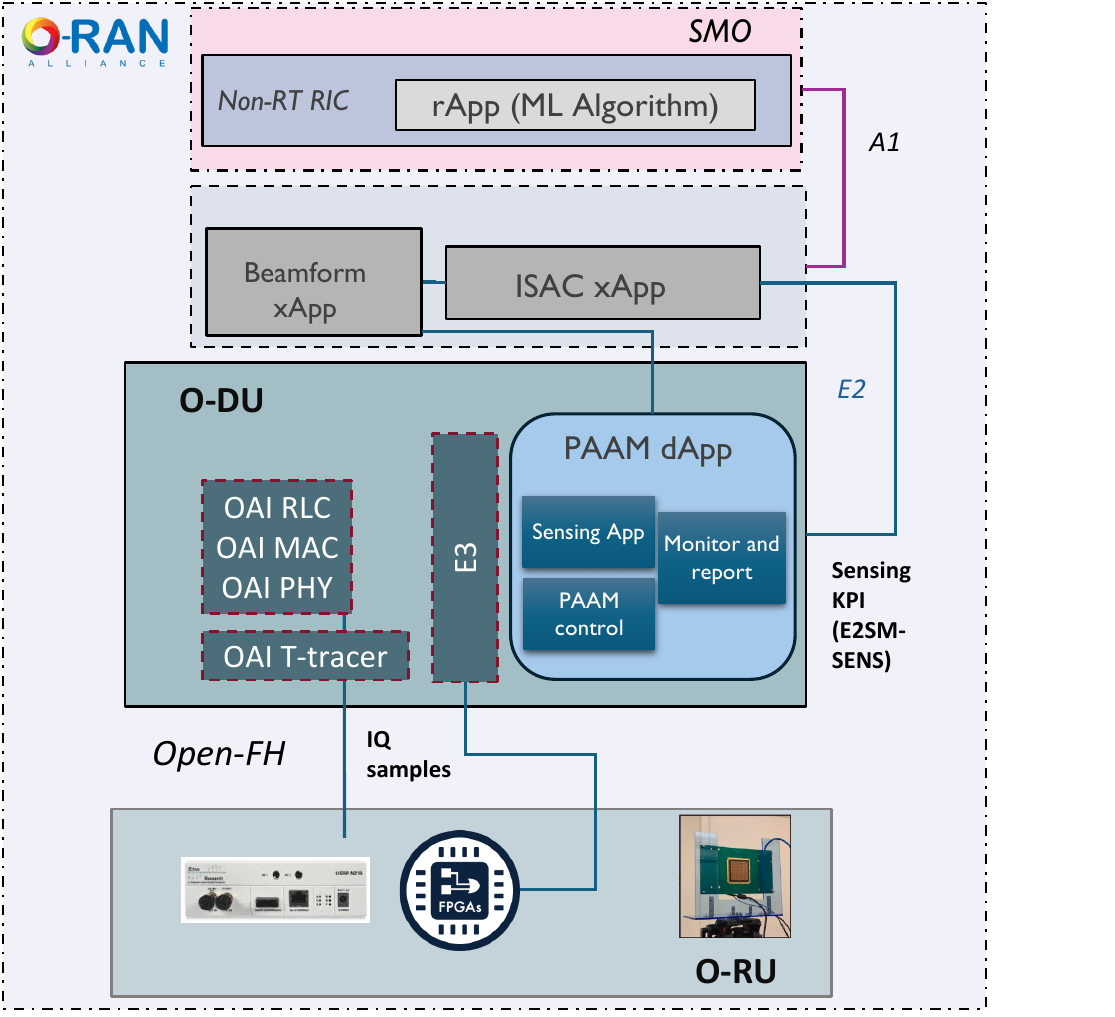}
\caption{PAAM dApp coordination with FD-enabled RU under the ISAC xApp.}
\label{fig:fd_beam_coordination}
\end{figure}

Fig.~\ref{fig:fd_beam_coordination} illustrates the coordination mechanism. The RU and DU exchange IQ samples via O-FH. The PAAM dApp processes these samples and reports metrics (Doppler, AoA, residual SI) via E2SM-SENS. The ISAC xApp subscribes to these metrics and performs arbitration when resource contention is detected. Using the localization telemetry received through E2SM-SENS, the xApp can employ learning-based approaches such as graph neural networks with discrete reparameterizations to jointly optimize beam allocation between sensing and communication under integer constraints. At the Non-RT RIC, rApps define long-term policies via A1, such as spatial sensing priorities (e.g., prioritizing vehicular corridors during peak hours) or adjusting the sensing-to-communication resource ratio based on historical traffic patterns and predicted demand.

\section{Implementation and Evaluation}
\label{sec:evaluation}

To validate the feasibility of the proposed architecture, we implemented a feasibility prototype of the E2SM-SENS telemetry pipeline and measured its end-to-end latency.

\subsection{Prototype Implementation}


The prototype implements the dApp-xApp interaction using FlexRIC~\cite{flexric2023}, an open-source O-RAN-compliant platform, with ZeroMQ as the inter-process transport between dApp and xApp. The sensing dApp processes IQ samples from USRP software-defined radios and generates telemetry reports containing spectral features (FFT-based). While the current implementation does not perform actual radar sensing (e.g., target detection or localization), it validates the latency characteristics of the E2SM-SENS pipeline that would transport such measurements in a production deployment. Each report carries a timestamp $t_0$ recorded at generation time. Upon reception, the xApp records timestamp $t_1$. Control commands from xApp to dApp are instrumented separately.

This instrumentation enables measurement of:
\begin{itemize}
    \item \textbf{Telemetry latency} ($t_1 - t_0$): Time from report generation at dApp to reception at xApp.
    \item \textbf{Control latency}: Time for a control command to propagate from xApp to dApp.
    \item \textbf{Closed-loop latency}: Sum of telemetry and control latencies, representing the minimum time for sensing feedback to trigger a control action.
\end{itemize}
\subsection{Experimental Validation}

We conducted two experiments across 10 trials with approximately 35,000 total samples.

\begin{figure*}[t]
\centering
\includegraphics[width=0.95\textwidth]{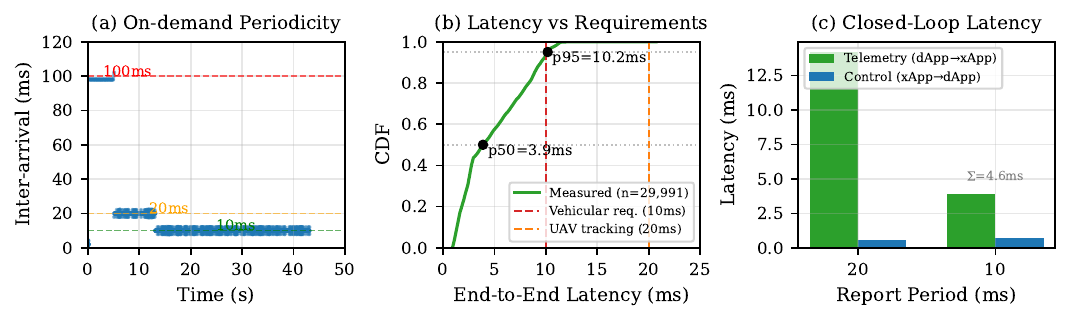}
\caption{Experimental validation (~35k samples): (a) on-demand periodicity control showing 100$\rightarrow$20$\rightarrow$10~ms transitions; (b) telemetry latency ($t_1 - t_0$) CDF at 10~ms period vs. use case thresholds; (c) closed-loop latency breakdown showing telemetry and control components.}
\label{fig:results}
\end{figure*}

\textbf{Experiment A (Periodicity Control):} The xApp dynamically adjusts the reporting period from 100$\rightarrow$20$\rightarrow$10~ms. Figure~\ref{fig:results}(a) shows the dApp tracks requested periods with mean inter-arrival within 0.1~ms of target. The p95 jitter at 10~ms is 8.4~ms, indicating that most reports arrive within the expected window.

\textbf{Experiment B (Closed-Loop Latency):} Figure~\ref{fig:results}(b) shows the telemetry latency CDF at 10~ms reporting period. The median is 3.9~ms (p95=10.2~ms), with 93.4\% of samples below the vehicular perception threshold ($<$10~ms) and 100\% below UAV tracking ($<$20~ms). Figure~\ref{fig:results}(c) decomposes the closed-loop latency: telemetry dominates (median 3.9~ms) while control latency adds minimal overhead (median 0.7~ms), yielding a total closed-loop median of 4.6~ms.

These results demonstrate that the E2SM-SENS architecture achieves closed-loop latencies compatible with vehicular perception and UAV tracking. Industrial control ($<$1~ms) would require optimizations such as kernel bypass or DPDK-based transport.

\section{Practical Deployment Considerations}
\label{sec:deployment}

The transition from prototype to production ISAC deployments raises several practical considerations that operators and vendors must address.

\textbf{Incremental adoption path:} The proposed architecture enables gradual ISAC integration without disrupting existing O-RAN deployments. Operators can initially deploy sensing dApps alongside communication functions, using E2SM-SENS subscriptions to collect environmental telemetry without modifying scheduler behavior. As confidence grows, xApps can progressively incorporate sensing feedback into resource allocation decisions. This staged approach reduces deployment risk while building operational experience.

\textbf{Hardware implications:} Monostatic sensing with Full-Duplex operation requires RU hardware capable of simultaneous transmit-receive with adequate self-interference cancellation. Modern phased array antenna modules increasingly embed FD capability that can be software-controlled, as demonstrated in our PAAM dApp implementation. The additional RF chain for the receive path during FD operation increases fronthaul bandwidth requirements by approximately 10--15\%, which remains within the capacity margins of typical Split~7.2 deployments~\cite{hamidi2024}.

\textbf{Spectrum and regulatory aspects:} ISAC operation reuses communication waveforms for sensing, avoiding dedicated radar spectrum allocations. However, the environmental data collected through sensing (object positions, movement patterns, occupancy information) raises privacy considerations that current regulatory frameworks do not fully address. Operators must implement data governance policies that comply with emerging guidelines while preserving the utility of sensing-derived insights.

\textbf{Multi-vendor interoperability:} The E2SM-SENS service model follows O-RAN's modular design philosophy, enabling sensing dApps from one vendor to interoperate with xApps from another through standardized telemetry schemas. This preserves the multi-vendor ecosystem that distinguishes O-RAN from traditional RAN architectures and allows operators to select best-of-breed components for their specific sensing requirements.

\section{Open Research Directions}
\label{sec:future}

Several research directions remain open for the community to address:

\textbf{GPU-accelerated sensing dApps:} The proposed architecture executes sensing algorithms on general-purpose DU hardware. Recent work on GPU-native gNBs~\cite{villa2025gpu} demonstrates that hardware acceleration can reduce inference latency to sub-2 ms for ML-based sensing pipelines. Integrating GPU acceleration into our monostatic sensing dApps would enable more sophisticated algorithms (neural network-based target classification, multi-target tracking, multi-modal fusion with camera or LiDAR inputs) while maintaining real-time constraints. The key challenge lies in efficient data movement between the O-FH receive path and GPU memory without introducing additional latency.

\textbf{dApp resource arbitration:} Multiple sensing and communication dApps may compete for shared RF chains, baseband processing cycles, or fronthaul bandwidth. Developing priority models, temporal slotting for sensing actions, or lightweight arbitration layers will be essential to ensure deterministic execution of time-critical ISAC tasks under dense deployment conditions.

\textbf{Radar processing placement:} A fundamental tradeoff exists between processing at the RU (reduced fronthaul load) versus the DU (COTS hardware, centralized security)~\cite{ludant2025isac_oran}. For massive-MIMO with Split~7.2x Category~B, RU-side preprocessing may be necessary, requiring standardized interfaces for compressed sensing features.

\textbf{Learning-based policy inference:} Designing rApps that process historical sensing KPIs to detect long-term environmental patterns or trigger spatially targeted sensing campaigns remains unexplored. Balancing sensing resolution against communication efficiency requires new policy models capturing application-level priorities across both domains.

\textbf{Multi-static coordination:} Extending to bistatic sensing requires sub-nanosecond synchronization accuracy for meter-level positioning. GPS-free synchronization through bistatic signal matching remains an open challenge. The E2SM-SENS foundation enables coordinated subscriptions across multiple DUs.

\textbf{Security and privacy:} Fronthaul IQ samples expose environmental information vulnerable to eavesdropping. Zero-trust frameworks and RU-side scrambling are needed~\cite{ludant2025isac_oran}. Sensing-derived location data requires anonymization techniques as regulatory frameworks evolve.

\textbf{Validation at scale:} Large-scale experimentation on testbeds such as COSMOS~\cite{raychaudhuri:mobicom2020} and x5G~\cite{x5g} is needed. Recent results show FR1 deployments achieving approximately 1~m RMSE localization, while cell-free massive MIMO with RIS assistance reaches sub-centimeter precision under controlled conditions.



\section{Conclusions}
\label{sec:conclusions}

This work presents architectural extensions enabling monostatic ISAC within O-RAN, addressing a critical gap as the industry moves toward 6G. We specify sensing dApps for real-time IQ processing at the O-DU, define E2SM-SENS for structured sensing telemetry with publisher-subscriber semantics, and identify O-FH metadata requirements for waveform-echo correlation. The architecture leverages Full-Duplex capability embedded in modern phased array hardware, controlled through a unified PAAM dApp that manages both beamforming and simultaneous transmit-receive operation.

Prototype evaluation demonstrates that the proposed architecture achieves closed-loop latencies compatible with vehicular perception and UAV tracking use cases. The experimental results reveal that E2 transport efficiency represents the primary optimization target for further latency reduction.

The proposed extensions maintain compatibility with O-RAN's hierarchical control structure while enabling sensing integration. Operators can adopt ISAC incrementally, starting with passive telemetry collection and progressively incorporating sensing feedback into resource allocation as operational confidence grows.

While focused on monostatic configurations, the E2SM-SENS interface provides a foundation for bistatic and cooperative sensing through coordinated multi-DU subscriptions. The standardized telemetry schemas preserve O-RAN's multi-vendor interoperability, allowing operators to combine sensing dApps and xApps from different vendors. As 6G standardization advances, the architectural patterns demonstrated here, including hierarchical control separation, structured telemetry exposure, and metadata-enriched fronthaul, offer a practical path toward production ISAC deployments.

\bibliographystyle{IEEEtran}
\bibliography{biblio}

\vskip -2\baselineskip plus -1fil

\begin{IEEEbiographynophoto}{Eduardo Baena} is a Postdoctoral Research Fellow at Northeastern University’s Institute for Intelligent Networked Systems, with research interests in AI-driven cellular network management, 5G/6G, Open RAN, cloud/edge computing, and NTN/LEO systems.\end{IEEEbiographynophoto}

\vskip -3\baselineskip plus -1fil

\begin{IEEEbiographynophoto}{Rajesh Krishnan} is a M.S. student at Northeastern University, with research interests in wireless systems, software-defined radios, and next-generation network architectures.\end{IEEEbiographynophoto}
\vskip -3\baselineskip plus -1fil
\begin{IEEEbiographynophoto}{Mai Vu} is a Professor of Electrical and Computer Engineering at Tufts University, where she leads LiNKS and works on wireless systems, signal processing, and 5G-and-beyond communications including ML and mmWave.\end{IEEEbiographynophoto}
\vskip -3\baselineskip plus -1fil
\begin{IEEEbiographynophoto}{Gil Zussman} is a Professor and Chair of Electrical Engineering at Columbia University, with research interests in wireless and resilient networks, and he leads the NSF-funded COSMOS testbed.\end{IEEEbiographynophoto}
\vskip -3\baselineskip plus -1fil
\begin{IEEEbiographynophoto}{Dimitrios Koutsonikolas} is an Associate Professor of Electrical and Computer Engineering at Northeastern University, with research interests in experimental wireless networking and mobile computing for 5G/NextG systems.\end{IEEEbiographynophoto}

\end{document}